\documentclass[notitlepage,a4paper,aps,prd,onecolumn,superscriptaddress,nofootinbib,groupedaddress,longbibliography]{revtex4-1}
\usepackage{amsmath}
\usepackage{amsfonts}
\usepackage{amssymb}
\usepackage{accents}
\usepackage[utf8]{inputenc}
\usepackage[T1]{fontenc}
\usepackage[colorlinks=true]{hyperref}
\usepackage{subfigure}
\usepackage{graphicx}

\newcommand{\dd}{\mathrm{d}}
\newcommand{\lc}[1]{\accentset{\circ}{#1}}
\newcommand{\vbr}[1]{\accentset{\mathcal{V}}{#1}}
\newcommand{\abr}[1]{\accentset{\mathcal{A}}{#1}}
\DeclareMathOperator{\sgn}{sgn}

\begin{document}

\title{Teleparallel axions and cosmology}

\author{Manuel Hohmann}
\email{manuel.hohmann@ut.ee}
\affiliation{Laboratory of Theoretical Physics, Institute of Physics, University of Tartu, W. Ostwaldi 1, 50411 Tartu, Estonia}

\author{Christian Pfeifer}
\email{christian.pfeifer@zarm.uni-bremen.de}
\affiliation{Laboratory of Theoretical Physics, Institute of Physics, University of Tartu, W. Ostwaldi 1, 50411 Tartu, Estonia}
\affiliation{Center Of Applied Space Technology And Microgravity - ZARM, University of Bremen, Am Fallturm 2, 28359 Bremen, Germany}

\begin{abstract}
We consider the most general teleparallel theory of gravity whose action is a linear combination of the five scalar invariants which are quadratic in the torsion tensor. Since two of these invariants possess odd parity, they naturally allow for a coupling to pseudo-scalar fields, thus yielding a Lagrangian which is even under parity transformations. In analogy to similar fields in gauge theories, we call these pseudo-scalar fields \emph{teleparallel axions}. For the most general coupling of a single axion field, we derive the cosmological field equations. We find that for a family of cosmologically symmetric teleparallel geometries, which possess non-vanishing axial torsion, the axion coupling contributes to the cosmological dynamics in the early universe. Most remarkably, this contribution is also present when the axion is coupled to the teleparallel equivalent of general relativity, hence allowing for a canonical coupling of a pseudo-scalar to general relativity. For this case we schematically present the influence of the axion coupling on the fixed points in the cosmological dynamics understood as dynamical system. Finally, we display possible generalizations and similar extensions in other geometric frameworks to model gravity.
\end{abstract}

\maketitle


\section{Introduction}\label{sec:intro}
Scalar fields which change their sign under parity transformations, so called pseudo-scalar fields, have been considered in physics since long. Most notably, they have been introduced to solve the strong CP violation problem in QCD~\cite{Peccei:1977hh,Wilczek:1977pj,Weinberg:1977ma}, where the name ``axion'' has been coined for the newly introduced pseudo-scalar field. Apart from QCD, it has been shown that they appear, both as a theoretical concept~\cite{Itin:2007cv,Hehl:2016yle} and experimentally verified in media~\cite{Obukhov:2005kh,Hehl:2007jy}, naturally in the axiomatic approach to general linear electrodynamics~\cite{Hehl:2003}. In the context of gravitational physics, axions are prominent candidates as dark matter particles~\cite{Duffy:2009ig,Sikivie:2009fv} and inflaton fields~\cite{Pajer:2013fsa,Marsh:2015xka}. In the context of scalar tensor extensions of general relativity they can be coupled to gravity, for example dynamically to the Chern-Simons term~\cite{Alexander:2009tp,Jackiw:2003pm}, or non-minimally in the context of $f(R)$-Gravity~\cite{Odintsov:2019evb}.

Besides the most well-known formulation of general relativity in terms of curvature, alternative formulations based on torsion or nonmetricity have been conceived~\cite{BeltranJimenez:2019tjy}, giving rise to the so called symmetric teleparallel equivalent of general relativity (STEGR)~\cite{Nester:1998mp} and the teleparallel equivalent of general relativity (TEGR)~\cite{Einstein:1928,Moller:1961,Maluf:2013gaa}. Here we will focus on the latter, and its extensions, known as teleparallel gravity~\cite{Aldrovandi:2013wha}. In these theories one considers a tetrad and a flat, metric compatible spin connection as fundamental variables, which encode the gravitational interaction in the torsion of a flat, metric-compatible connection, in contrast to the metric and its torsion free Levi-Civita connection, which describes gravity in terms of curvature. The advantage in the teleparallel approach is that it can be interpreted as a gauge theory of gravity~\cite{Hayashi:1967se,Cho:1975dh,Hayashi:1977jd} and that it allows for numerous extension of general relativity without introducing higher than second order derivative field equations~\cite{Ferraro:2006jd,Ferraro:2008ey,Linder:2010py,Maluf:2011kf,Bahamonde:2017wwk,Hohmann:2017duq}. Modifying teleparallel theories of gravity with additional non-minimally coupled scalar fields has been studied throughout the literature~\cite{Geng:2011aj,Bamba:2013jqa,Bahamonde:2015hza,Hohmann:2018rwf,Hohmann:2018vle,Hohmann:2018dqh,Hohmann:2018ijr,Bahamonde:2019shr}, and we will extend this class in this article by considering a non-minimal coupling to a pseudo-scalar field. This vast variety of possible teleparallel theories of gravity is possible since their building block is the torsion tensor of the flat connection, which contains only first derivatives of the tetrad. Thus, an arbitrary high number of products of torsion tensors can be considered in the action and still the field equations will be of second order.

Focusing on the teleparallel theories of gravity built from actions which are quadratic in the torsion tensor, there exist five independent scalar torsion invariants~\cite{Hayashi:1979qx,Hayashi:1981qx,Bahamonde:2017wwk}, three of which possess even parity, while the remaining two are of odd parity. The most general quadratic teleparallel theory of gravity built from the parity-even invariants, which can be called the axial, vectorial and tensorial torsion scalar, is known as new general relativity (NGR)~\cite{Hayashi:1979qx,Hayashi:1981qx}, of which TEGR is a special case. The parity-odd terms, however, have gained less attention in the literature so far. Recently they have been considered to construct a parity violating extension of TEGR~\cite{Li:2020xjt,Chatzistavrakidis:2020wum}, in which a pseudo-scalar field is coupled to the Nieh-Yan term~\cite{Nieh:1981ww}.

In this article, we generalize the theory presented in~\cite{Li:2020xjt,Chatzistavrakidis:2020wum} and construct the most general quadratic teleparallel theory of gravity employing all five quadratic torsion scalar terms. In particular, we use the two parity-odd torsion to naturally couple a pseudo-scalar field to gravity, which we call \emph{teleparallel axion}. A particularly interesting subclass emerges when we reduce the part of the action constructed from the three parity-even torsion invariants to TEGR. In this way the teleparallel torsion based formulation of general relativity allows for a natural coupling of pseudo-scalars. Both for the general theory and the special case of the axion coupling to TEGR, we study the cosmological dynamics. Moreover, we propose a number of further generalized theories, which one may expect to exhibit similar properties as the one we study in this article.

This article is structured as follows. In section~\ref{sec:telegeom} we state the basics of teleparallel geometry, and list the scalar invariants which can be constructed within this geometry. The action and field equations for our model are shown in section~\ref{sec:action}. Its cosmological dynamics are derived in section~\ref{sec:cosmo}. Further possibilities to extend and modify this model are discussed in section~\ref{sec:ext}. We end with a summary and outlook in section~\ref{sec:conclusion}.

The notation conventions throughout this article are that lowercase Greek indices \(\mu, \nu, \ldots\) run from \(0\) to \(3\) label spacetime coordinate bases, while lowercase Latin indices \(a, b, \ldots\) label Lorentz bases, and likewise run from \(0\) to \(3\). Capital Latin indices \(A, B, \ldots\) label scalar field components.

\section{Teleparallel Geometry}\label{sec:telegeom}
Before we present the general coupling of axions to teleparallel theories of gravity, whose action is composed from terms which are quadratic in the torsion tensor, we briefly review the necessary mathematical notions and the constituent terms for these theories, thereby declaring the conventions we will be using in this article. The fundamental geometric notions of teleparallel geometry are summarized in section~\ref{ssec:tpgeom}. We then list the quadratic torsion invariants, which can be defined in this geometrical framework, in section~\ref{ssec:teleinv}.

\subsection{The general setting}\label{ssec:tpgeom}
The fundamental field variables in the covariant formulation of teleparallel gravity are the tetrad \(\theta^a{}_{\mu}\) and a flat Lorentz spin connection \(\omega^a{}_{b\mu}\)~\cite{Aldrovandi:2013wha,Krssak:2015oua}. The flatness and metric compatibility demand on the spin connection imply that locally it can be written in the form
\begin{equation}\label{eq:scon}
\omega^a{}_{b\mu} = \Lambda^a{}_c \partial_\mu (\Lambda^{-1})^c{}_b\,,
\end{equation}
where $\Lambda^a{}_c$ are the components of a Lorentz transformation satisfying
\begin{equation}
\eta_{ab}\Lambda^a{}_c\Lambda^b{}_d = \eta_{cd}\,,
\end{equation}
and $\eta_{ab} = \mathrm{diag}(-1,1,1,1)$ is the Minkowski metric. With the help of the latter, the tetrad further defines the metric
\begin{equation}
g_{\mu\nu} = \eta_{ab}\theta^a{}_{\mu}\theta^b{}_{\nu}\,.
\end{equation}
For the tetrad, one demands the existence of an inverse \(e_a{}^{\mu}\), which satisfies
\begin{equation}
\theta^a{}_{\mu}e_b{}^{\mu} = \delta^a_b\,, \quad
\theta^a{}_{\mu}e_a{}^{\nu} = \delta_{\mu}^{\nu}\,.
\end{equation}
Using the inverse, one can define the components of an affine connection, given by
\begin{equation}
\Gamma^{\mu}{}_{\nu\rho} = e_a{}^{\mu}(\partial_{\rho}\theta^a{}_{\nu} + \omega^a{}_{b\rho}\theta^b{}_{\nu})\,.
\end{equation}
It follows from the structure~\eqref{eq:scon} that this connection has vanishing curvature and nonmetricity,
\begin{equation}\label{eq:telep}
R^{\rho}{}_{\sigma\mu\nu} = \partial_{\mu}\Gamma^{\rho}{}_{\sigma\nu} - \partial_{\nu}\Gamma^{\rho}{}_{\sigma\mu} + \Gamma^{\rho}{}_{\lambda\mu}\Gamma^{\lambda}{}_{\sigma\nu} - \Gamma^{\rho}{}_{\lambda\nu}\Gamma^{\lambda}{}_{\sigma\mu} \equiv 0\,, \quad
Q_{\rho\mu\nu} = \nabla_{\rho}g_{\mu\nu} \equiv 0\,.
\end{equation}
Its torsion
\begin{equation}\label{eq:torsion}
T^{\rho}{}_{\mu\nu} = \Gamma^{\rho}{}_{\nu\mu} - \Gamma^{\rho}{}_{\mu\nu}\,,
\end{equation}
however, is in general non-vanishing. These properties distinguish the teleparallel connection \(\Gamma^{\mu}{}_{\nu\rho}\) from the Levi-Civita connection \(\lc{\Gamma}^{\mu}{}_{\nu\rho}\), which has vanishing torsion, bu non-vanishing curvature, and whose components, as well as related quantities, we denote with an empty circle. The difference between the coefficients of the two connections is measured by the contortion
\begin{equation}\label{eq:contor}
K^{\mu}{}_{\nu\rho} = \Gamma^{\mu}{}_{\nu\rho} - \lc{\Gamma}^{\mu}{}_{\nu\rho} = \frac{1}{2}\left(T_{\nu}{}^{\mu}{}_{\rho} + T_{\rho}{}^{\mu}{}_{\nu} - T^{\mu}{}_{\nu\rho}\right)\,.
\end{equation}
Due to the form~\eqref{eq:scon} of the connection $\omega^a{}_{b\mu}$ it is always possible to make a Lorentz transformation to the so called Weitzenböck gauge, in which the tetrad is transformed to $\tilde{\theta}^a{}_{\mu} = \theta^b{}_{\mu}(\Lambda^{-1})^a{}_b$ and the spin connection $\tilde{\omega}^a{}_{b\mu}{} = 0$ vanishes. In this gauge, the torsion takes the simple form\\
\begin{align}
T^{\rho}{}_{\mu\nu} = \tilde{e}_a{}^{\rho}(\partial_{\mu}\tilde{\theta}^a{}_{\nu} - \partial_{\nu}\tilde{\theta}^a{}_{\mu})\,.
\end{align}
We will work in the this gauge when we display the cosmologically symmetric tetrad in section~\ref{ssec:cosmotet} and drop the tilde for brevity. Further, we use the convention that spacetime (Greek) and Lorentz (Latin) indices can be changed into each other by contraction with the tetrad and its inverse.

\subsection{Scalar and pseudo-scalar torsion invariants}\label{ssec:teleinv}
The torsion tensor~\eqref{eq:torsion} is obtained as the exterior covariant derivative of the tetrad one-forms with respect to the spin connection, and can be interpreted as a gauge field strength in the gauge theory picture of teleparallel gravity~\cite{Aldrovandi:2013wha}. Hence, a canonical choice for the kinetic energy term for a tetrad in a teleparallel gravity action is to consider scalar invariants which are quadratic in the torsion tensor. Due to the symmetries of the torsion, there exist five independent scalars and pseudo-scalars, which are quadratic in the torsion~\cite{Hehl:1994ue}. This can be seen by decomposing the torsion tensor in a vector, axial and tensor part given by
\begin{align}
T^{\mu}{}_{\nu\rho} = \mathfrak{V}^{\mu}{}_{\nu\rho} + \mathfrak{A}^{\mu}{}_{\nu\rho} + \mathfrak{T}^{\mu}{}_{\nu\rho}\,.
\end{align}
The different contributions to the torsion are defined in the terms of the quantities
\begin{subequations}
\begin{align}
\mathfrak{v}_{\mu} &= T^{\nu}{}_{\nu\mu}\,, &
\mathfrak{V}^{\mu}{}_{\nu\rho} &= \frac{2}{3}\delta^{\mu}_{[\nu}\mathfrak{v}_{\rho]}\,,\label{eq:vecT}\\
\mathfrak{a}_{\mu} &= \frac{1}{6}\epsilon_{\mu\nu\rho\sigma}T^{\nu\rho\sigma}\,, &
\mathfrak{A}_{\mu\nu\rho} &= \epsilon_{\mu\nu\rho\sigma}\mathfrak{a}^{\sigma}\,,\label{eq:axT}\\
\mathfrak{t}_{\mu\nu\rho} &= T_{(\mu\nu)\rho} + \frac{1}{3}\left(T^{\sigma}{}_{\sigma(\mu}g_{\nu)\rho} - T^{\sigma}{}_{\sigma\rho}g_{\mu\nu}\right)\,, &
\mathfrak{T}^{\mu}{}_{\nu\rho} &= \frac{4}{3}\mathfrak{t}^{\mu}{}_{[\nu\rho]}\,,\label{eq:tenT}
\end{align}
\end{subequations}
which satisfy the identities
\begin{align}
\mathfrak{A}^{\nu}{}_{\nu\mu} = \mathfrak{T}^{\nu}{}_{\nu\mu} = 0\,, \quad \mathfrak{V}_{[\mu\nu\rho]} = \mathfrak{T}_{[\mu\nu\rho]} = 0\,.
\end{align}
These can also be used to decompose the contortion~\eqref{eq:contor} in the form
\begin{equation}\label{eq:contoratv}
K_{\mu\nu\rho} = \frac{1}{2}\epsilon_{\mu\nu\rho\sigma}\mathfrak{a}^{\sigma} + \frac{2}{3}\mathfrak{v}_{[\mu}g_{\nu]\rho} - \frac{4}{3}\mathfrak{t}_{\rho[\mu\nu]}\,,
\end{equation}
which we note here for later use. There are three independent parity even quadratic scalars that can be constructed
\begin{align}\label{eq:tseven}
T_{\text{vec}} = \mathfrak{v}_{\mu}{\mathfrak{v}}^{\mu}\,, \quad
T_{\text{axi}} = \mathfrak{a}_{\mu}{\mathfrak{a}}^{\mu}\,, \quad
T_{\text{ten}} = \mathfrak{t}_{\lambda\mu\nu}{\mathfrak{t}}^{\lambda\mu\nu}\,,
\end{align}
and two independent real parity odd ones
\begin{align}\label{eq:tsodd}
P = \mathfrak{v}_{\mu}\mathfrak{a}^{\mu}, \quad
\tilde{P} = \epsilon_{\mu\nu\rho\sigma}\mathfrak{t}_{\lambda}{}^{\mu\nu}\mathfrak{t}^{\lambda\rho\sigma}\,.
\end{align}
Constructing an action from an linear combination of the parity-even terms~\eqref{eq:tseven} one thus arrives at a teleparallel gravity theory which is invariant under parity transformation~\cite{Hayashi:1979qx,Hayashi:1981qx}. Including the parity-odd terms in the same fashion, however, leads to parity violation. This parity transformation can naturally be absorbed by coupling them to a pseudo-scalar field, which likewise changes its sign under parity transformation. In the next section we derive the field equations for the most general action of this type, which is linear in the five torsion invariants and contains a pseudo-scalar field coupling for the parity-odd terms.

\section{Action and field equations}\label{sec:action}
From the torsion scalars introduced in the previous section we seek to construct the most general action which is linear in these scalar invariants and even under parity transformations. In order to do so we introduce a dynamical pseudo-scalar field $\phi$ which serve as dynamical coupling constant to the parity odd terms in section~\ref{ssec:action}. After the construction of the action we derive the corresponding field equations~\ref{ssec:feq}. A few special cases and related theories are pointed out in section~\ref{ssec:related}.

\subsection{The Action}\label{ssec:action}
In the following, we will assume that the fundamental dynamical fields we consider are the tetrad \(\theta^a{}_{\mu}\), the spin connection \(\omega^a{}_{b\mu}\), a pseudo-scalar field \(\phi\), as well as a set \(\chi\) of matter fields, which we do not specify further. Their dynamics is chosen to be governed by an action of the form
\begin{align}\label{eq:action}
S[\theta, \omega, \phi, \chi] = S_{\text{g}}[\theta, \omega, \phi] + S_{\text{m}}[\theta, \chi]\,,
\end{align}
and hence splits into a gravitational part \(S_{\text{g}}\) and a matter part \(S_{\text{m}}\). For the latter we assume that it only depends on the matter fields \(\chi\) and the tetrad \(\theta^a{}_{\mu}\), but not on the teleparallel spin connection \(\omega^a{}_{b\mu}\). This assumption is made purely for simplicity, as it will not be relevant for the axion extension of the gravitational action  \(S_{\text{g}}\) we discuss here. One could also choose to couple the matter to the teleparallel spin connection, a thorough discussion of these options can for example be found in~\cite{BeltranJimenez:2020sih}. However, for the cosmological analysis we perform in Section~\ref{sec:cosmo}, both couplings lead to the same results, and so we will restrict ourselves to the simpler choice.

It follows from the aforementioned assumption on the matter coupling that we can write the variation of the matter action \(S_{\text{m}}\) with respect to the tetrad in the form
\begin{equation}
\delta_{\theta}S_{\text{m}} = -\int_M\dd^4x\theta\,\Theta_a{}^{\mu}\,\delta\theta^a{}_{\mu}\,,
\end{equation}
where \(\Theta_a{}^{\mu}\) denotes the energy-momentum tensor. Further assuming that the matter action is invariant under local Lorentz transformations, we find that the energy-momentum tensor must be symmetric~\cite{Hohmann:2017duq}, hence \(\Theta_{[\mu\nu]} = 0\). Note that in general this would not be the case if we had included a coupling to the spin connection, which would also enter the condition of local Lorentz invariance, and another, antisymmetric contribution related to hypermomentum would arise, which has the same origin as the corresponding term obtained from the variation of the gravitational part of the action~\cite{Hohmann:2021fpr}. However, for the special case of cosmological symmetry, which is the case we investigate in the Section~\ref{sec:cosmo} in more detail, the choice of matter coupling does not make any difference, since, due to the spacetime symmetry conditions the field equations obey the contribution of the hypermomentum to the field equations vanishes~\cite{Hohmann:2019nat}. For the gravitational part \(S_{\text{g}}\) of the action, we assume the form
\begin{equation}\label{eq:gravact}
S_{\text{g}}[\theta, \omega, \phi] = \int\dd^4x\,\theta\,\frac{1}{2\kappa^2}\left[c_vT_{\text{vec}} + c_aT_{\text{axi}} + c_tT_{\text{ten}} + b\phi P + \tilde{b}\phi\tilde{P} + \mathcal{Z}(\phi)g^{\mu\nu}\partial_{\mu}\phi\partial_{\nu}\phi + 2\kappa^2\mathcal{V}(\phi)\right]\,,
\end{equation}
and which is the most general Lagrangian which satisfies the following restrictive assumptions:
\begin{enumerate}
\item
The three terms \(T_{\text{vec}}\), \(T_{\text{axi}}\), \(T_{\text{ten}}\) are the only possible scalars which are quadratic in the torsion tensor and of even parity. Here \(c_v, c_a, c_t\) are free constants, so that this part of the action resembles new general relativity~\cite{Hayashi:1979qx,Hayashi:1981qx}. If these constants take the values
\begin{equation}\label{eq:tegr}
c_a = \frac{3}{2}\,, \quad
c_v = -\frac{2}{3}\,, \quad
c_t = \frac{2}{3}\,,
\end{equation}
then this part of the action reduces to TEGR.
\item
The two terms \(P\) and \(\tilde{P}\) are the remaining terms which are quadratic in the torsion tensor, but these have odd parity. In order to obtain parity-even terms in the action, we multiply each terms with one of the pseudo-scalar field \(\phi\), introducing two further free coupling constants \(b\) and \(\tilde{b}\).
\item
The remaining terms are the kinetic term depending on a kinetic coupling function \(\mathcal{Z}\), as well as a potential \(\mathcal{V}\). These functions are assumed to be even in their arguments, so that the resulting terms are of even parity.
\end{enumerate}
The role of the two pseudo-scalar field \(\phi\), appearing non-minimally coupled to the parity-odd, quadratic torsion terms \(P\) and \(\tilde{P}\), is reminiscent of the QCD axion field in the Peccei-Quinn theory~\cite{Peccei:1977hh}, or the abelian axion field in general linear electrodynamics~\cite{Hehl:2003}. Hence, it is justified to call this field a \emph{teleparallel axion}.

\subsection{Field equations}\label{ssec:feq}
Variation of the action with respect to the tetrad yields
\begin{equation}\label{eq:tetradfeq}
\begin{split}
2\kappa^2\Theta_{\mu\nu} &= c_a\left[\mathfrak{a}^{\rho}\mathfrak{a}_{(\rho}g_{\mu\nu)} - \frac{8}{9}\epsilon_{\nu\alpha\beta\gamma}\mathfrak{a}^{\alpha}\mathfrak{t}_{\mu}{}^{\beta\gamma} - \frac{4}{9}\epsilon_{\mu\nu\rho\sigma}\mathfrak{a}^{\rho}\mathfrak{v}^{\sigma} - \frac{2}{3}\epsilon_{\mu\nu\rho\sigma}\lc{\nabla}^{\rho}\mathfrak{a}^{\sigma}\right]\\
&\phantom{=}+ c_v\left[\mathfrak{v}^{\rho}\mathfrak{v}_{(\rho}g_{\mu\nu)} + \frac{8}{3}\mathfrak{t}_{\mu[\rho\nu]}\mathfrak{v}^{\rho} + 4g_{\mu[\nu}\lc{\nabla}^{\rho}\mathfrak{v}_{\rho]} - \epsilon_{\mu\nu\rho\sigma}\mathfrak{a}^{\rho}\mathfrak{v}^{\sigma}\right]\\
&\phantom{=}+ c_t\left[\frac{4}{3}\mathfrak{t}_{\alpha[\beta\gamma]}\mathfrak{t}^{\alpha\beta\gamma}g_{\mu\nu} - \frac{8}{3}\mathfrak{t}_{\mu[\rho\sigma]}\mathfrak{t}_{\nu}{}^{\rho\sigma} + 4\lc{\nabla}^{\rho}\mathfrak{t}_{\mu[\nu\rho]} - \frac{4}{3}\mathfrak{t}_{\nu[\mu\rho]}\mathfrak{v}^{\rho} + \epsilon_{\mu\alpha\beta\gamma}\mathfrak{a}^{\alpha}\mathfrak{t}_{\nu}{}^{\beta\gamma}\right]\\
&\phantom{=}+ b\phi\left[\mathfrak{a}^{\rho}\mathfrak{v}_{(\rho}g_{\mu\nu)} + 2\frac{g_{\mu[\nu}\lc{\nabla}^{\rho}(\phi\mathfrak{a}_{\rho]})}{\phi} - \frac{1}{3}\epsilon_{\mu\nu\rho\sigma}\frac{\lc{\nabla}^{\rho}(\phi\mathfrak{v}^{\sigma})}{\phi} - \frac{4}{3}\mathfrak{t}_{\mu[\nu\rho]}\mathfrak{a}^{\rho} - \frac{4}{9}\epsilon_{\nu\alpha\beta\gamma}\mathfrak{v}^{\alpha}\mathfrak{t}_{\mu}{}^{\beta\gamma}\right]\\
&\phantom{=}+ \tilde{b}\phi\bigg[6g_{\nu[\mu}\epsilon_{\alpha\beta]\gamma\delta}\mathfrak{t}^{\rho\alpha\beta}\mathfrak{t}_{\rho}{}^{\gamma\delta} - \frac{8}{3}\mathfrak{t}_{\mu}{}^{[\alpha\beta]}\mathfrak{t}_{(\nu}{}^{\gamma\delta}\epsilon_{\alpha)\beta\gamma\delta} - \frac{4}{3}g_{\nu\rho}g_{\mu[\sigma}\mathfrak{v}_{\alpha]}\mathfrak{t}^{(\rho}{}_{\beta\gamma}\epsilon^{\sigma)\alpha\beta\gamma} - 3\mathfrak{t}_{\nu[\mu\rho]}\mathfrak{a}^{\rho}\\
&\phantom{=}- 4\frac{g_{\mu\rho}g_{\nu[\sigma}\epsilon^{\alpha\beta\gamma(\rho}\lc{\nabla}_{\alpha]}(\phi\mathfrak{t}^{\sigma)}{}_{\beta\gamma})}{\phi}\bigg] - 2\mathcal{Z}\partial_{\mu}\phi\partial_{\nu}\phi + \mathcal{Z}g_{\mu\nu}g^{\rho\sigma}\partial_{\rho}\phi\partial_{\sigma}\phi + 2\kappa^2\mathcal{V}g_{\mu\nu}\,,
\end{split}
\end{equation}
while variation with respect to the axion yields the generalized Klein-Gordon equation
\begin{equation}\label{eq:scalarfeq}
0 = \mathcal{Z}\lc{\square}\phi + \frac{1}{2}g^{\mu\nu}\mathcal{Z}_{\phi}\partial_{\mu}\phi\partial_{\nu}\phi - \kappa^2\mathcal{V}_{\phi} - bP - \tilde{b}\tilde{P}\,,
\end{equation}
where \(\mathcal{Z}_{\phi}\) and \(\mathcal{V}_{\phi}\) denote the derivatives of \(\mathcal{Z}\) and \(\mathcal{V}\) with respect to \(\phi\). We do not display the field equation arising from variation with respect to the spin connection, since it is identical to the antisymmetric part of the tetrad field equations~\cite{Golovnev:2017dox,Hohmann:2017duq}. Note that this antisymmetric equation has no matter contribution due to our assumption that the matter does not couple to the spin connection. However, this will not restrict the following analysis, in which we investigate these equations in cosmological symmetry, since any antisymmetric two-tensor with cosmological symmetry vanishes identically, as shown in~\cite{Hohmann:2019nat}, and therefore does not contribute to the cosmological dynamics.

\subsection{Special cases and related theories}\label{ssec:related}
For a general Riemann-Cartan connection, the curvature tensor \(R^{\mu}{}_{\nu\rho\sigma}\) is given in terms of the curvature tensor \(\lc{R}^{\mu}{}_{\nu\rho\sigma}\) of the Levi-Civita connection and the contortion \(K^{\mu}{}_{\nu\rho}\) as
\begin{equation}\label{eq:lccurv}
R^{\mu}{}_{\nu\rho\sigma} = \lc{R}^{\mu}{}_{\nu\rho\sigma} + \lc{\nabla}_{\rho}K^{\mu}{}_{\nu\sigma} - \lc{\nabla}_{\sigma}K^{\mu}{}_{\nu\rho} + K^{\mu}{}_{\tau\rho}K^{\tau}{}_{\nu\sigma} - K^{\mu}{}_{\tau\sigma}K^{\tau}{}_{\nu\rho}\,.
\end{equation}
Two parity-odd terms are readily obtained from this expression. These are the Holst term~\cite{Holst:1995pc}
\begin{equation}
\hat{R} = \frac{1}{2}\epsilon^{\mu\nu\rho\sigma}R_{\mu\nu\rho\sigma} = \epsilon^{\mu\nu\rho\sigma}\left(\lc{\nabla}_{\rho}K_{\mu\nu\sigma} + K_{\mu\tau\rho}K^{\tau}{}_{\nu\sigma}\right)
\end{equation}
and the topological Nieh-Yan term~\cite{Nieh:1981ww}
\begin{equation}
\tilde{R} = \lc{\nabla}_{\mu}\mathfrak{a}^{\mu} = \frac{1}{3}\epsilon^{\mu\nu\rho\sigma}\lc{\nabla}_{\mu}K_{\nu\rho\sigma} = \frac{1}{6}\epsilon^{\mu\nu\rho\sigma}(R_{\mu\nu\rho\sigma} - 2K_{\mu\tau\rho}K^{\tau}{}_{\nu\sigma})\,.
\end{equation}
Note that the Riemann tensor obtained from the Levi-Civita connection does not appear in these terms, since its totally antisymmetric part vanishes as a consequence of the first Bianchi identity; hence, no such parity violating terms may be constructed in Riemannian geometry, where the torsion tensor vanishes. In Riemann-Cartan geometry, however, coupling of scalar fields to these terms has been considered~\cite{Langvik:2020nrs}, and a coupling to pseudo-scalar fields is straightforward to conceive.

In teleparallel geometry, the curvature vanishes, \(R^{\mu}{}_{\nu\rho\sigma} \equiv 0\). Hence, also the Holst term vanishes identically. The Nieh-Yan term is expressed in terms quadratic in the contortion, and hence the torsion. Using the decomposition~\eqref{eq:contoratv}, it can be written in the form
\begin{equation}
\tilde{R} = -\frac{1}{3}\epsilon^{\mu\nu\rho\sigma}K_{\mu\tau\rho}K^{\tau}{}_{\nu\sigma} = \frac{2}{3}\mathfrak{v}_{\mu}\mathfrak{a}^{\mu} - \frac{4}{27}\epsilon_{\mu\nu\rho\sigma}\mathfrak{t}_{\lambda}{}^{\mu\nu}\mathfrak{t}^{\lambda\rho\sigma} = \frac{2}{3}P - \frac{4}{27}\tilde{P}\,,
\end{equation}
and thus becomes a linear combination of the two parity violating terms we introduced in the action~\eqref{eq:action}. In the particular case \(2b + 9\tilde{b} = 0\), the axion coupling term in the action~\eqref{eq:action} therefore reduces to a coupling to the Nieh-Yan term, and can equivalently be written as
\begin{equation}\label{eq:niehyan}
b\phi\left(P - \frac{2}{9}\tilde{P}\right) = \frac{3}{2}b\phi\tilde{R} = \frac{3}{2}b\phi\lc{\nabla}_{\mu}\mathfrak{a}^{\mu}\,,
\end{equation}
which can be integrated by parts, and is then equivalent to
\begin{equation}
-\frac{3}{2}b\mathfrak{a}^{\mu}\lc{\nabla}_{\mu}\phi\,,
\end{equation}
up to a boundary term. Hence, it takes a similar role to the coupling \(\mathfrak{v}^{\mu}\lc{\nabla}_{\mu}\phi\) of a scalar field to the vector torsion, which plays an important role in conformal transformations of scalar-torsion theories~\cite{Hohmann:2018dqh,Hohmann:2018ijr}, and which can equivalently be expressed in the form \(\phi\lc{\nabla}_{\mu}\mathfrak{v}^{\mu}\) as a dynamical coupling to the teleparallel boundary term \(B = \lc{\nabla}_{\mu}\mathfrak{v}^{\mu}\).

Finally, we highlight that the parity-odd teleparallel torsion invariants give a canonical way to couple axions to general relativity, which cannot be obtained from its more commonly used, curvature-based formulation. By choosing the constant parameters for the three parity-even torsion invariants to take their TEGR values~\eqref{eq:tegr}, the left hand side of the tetrad field equations \eqref{eq:tetradfeq} reduces to the Einstein tensor, which originates from the three parity-even torsion scalars in the action~\eqref{eq:gravact}, with an additive contribution from the axion field, i.e., axion and torsion induced modifications of the field equations of general relativity. Schematically, we can split the field equations into their symmetric and antisymmetric parts, which take the form
\begin{subequations}
\begin{align}
G_{\mu\nu} + \frac{1}{2}bP_{(\mu\nu)} + \frac{1}{2}\tilde{b}\tilde{P}_{(\mu\nu)} - \mathcal{Z}\partial_{\mu}\phi\partial_{\nu}\phi + \frac{1}{2}\mathcal{Z}g_{\mu\nu}g^{\rho\sigma}\partial_{\rho}\phi\partial_{\sigma}\phi + \kappa^2\mathcal{V}g_{\mu\nu} &= \kappa^2\Theta_{\mu\nu}\,,\\
\frac{1}{2}bP_{[\mu\nu]} + \frac{1}{2}\tilde{b}\tilde{P}_{[\mu\nu]} &= 0\,,
\end{align}
\end{subequations}
where \(P_{\mu\nu}\) and \(\tilde{P}_{\mu\nu}\) are the respective terms in square brackets in the field equations~\eqref{eq:tetradfeq}, which originate from the axion couplings. Note that these are the only terms which contribute to the antisymmetric part of the field equations, since the remaining contributions from the kinetic and potential parts of the axion are symmetric tensors, and this holds also for the Einstein and energy-momentum tensors. Our axion and torsion modified general relativity is a straightforward generalization of the theory presented in~\cite{Li:2020xjt,Chatzistavrakidis:2020wum}, where the axion is coupled solely to the Nieh-Yan term~\eqref{eq:niehyan}. The latter is obtained from the model we just presented by fixing the parameters \(2b + 9\tilde{b} = 0\).

\section{Cosmological dynamics}\label{sec:cosmo}
We now study the cosmological dynamics of the teleparallel axion model, whose action and field equations we displayed in the preceding section. The two branches of cosmologically symmetric tetrads we consider are shown in section~\ref{ssec:cosmotet}. These are used in section~\ref{ssec:cosmofeq} to derive the cosmological field equations. We finally take a closer look at their structure and the axion contribution to the cosmological dynamics in section~\ref{ssec:cosmoana}.

\subsection{Homogeneous and isotropic tetrad}\label{ssec:cosmotet}
The cosmological principle implies that the gravitational field of the universe as a whole must be described by a homogeneous and isotropic geometry. In the case of teleparallel geometry which we consider here, this symmetry condition implies that there exist precisely two canonical forms of the tetrad in the Weitzenböck, obtained from two different possibilities to represent the symmetry group of three dimensional homogeneous and isotropic spaces in the Lorentz group~\cite{Hohmann:2019nat,Hohmann:2020zre}. To display these standard forms, which we denote by $\vbr{\theta}^a{}_{\mu}$, called the vector branch, and $\abr{\theta}^a{}_{\mu}$, called the axial branch, we employ standard cosmological coordinates $(t,r,\vartheta,\varphi)$, a scale factor $A=A(t)$ and a lapse function $N=N(t)$. In terms of these, we can write the tetrads in matrix form as
\begin{align}\label{eq:tet1}
\vbr{\theta}^a{}_{\mu} =
\begin{pmatrix}
N \chi & i u A \frac{r}{\chi} & 0 & 0\\
i u N r \sin \vartheta \cos \varphi & A \sin \vartheta \cos \varphi  & A r \cos\vartheta \cos\varphi  & - A r \sin\vartheta \sin\varphi\\
i u N r \sin \vartheta \sin \varphi & A \sin \vartheta \sin \varphi  & A r \cos\vartheta \sin\varphi  &   A r \sin\vartheta \cos\varphi\\
i u N r \cos \vartheta & A \cos \vartheta & - A r \sin\vartheta  & 0
\end{pmatrix}\,,
\end{align}
and
\begin{align}\label{eq:tet2}
\abr{\theta}^a{}_{\mu} =
\begin{pmatrix}
N  & 0 & 0 & 0\\
0 & A \frac{\sin \vartheta \cos \varphi}{\chi}  & A r (\chi \cos\vartheta \cos\varphi + u r \sin\varphi)  & - A r \sin\vartheta ( \chi \sin\varphi - u r \cos\vartheta \cos\varphi )\\
0 & A \frac{\sin \vartheta \sin \varphi}{\chi}  & A r (\chi \cos\vartheta \sin\varphi + u r \cos\varphi)  &  A r \sin\vartheta ( \chi \cos\varphi + u r \cos\vartheta \sin\varphi )\\
0 & A \frac{\cos \vartheta}{\chi} & - A r \chi \sin\vartheta  & - u A r^2 \sin^2\vartheta
\end{pmatrix}\,,
\end{align}
where $\chi = \sqrt{1 - u^2r^2}$ and $k = u^2$ determines the sign of the curvature on the spatial hypersurfaces. Here \(u\) may be real or imaginary, so that \(k\) may be positive or negative. As a consequence, observe that one tetrad becomes complex, while the other one remains real, depending on the choice of the sign of the spatial curvature parameter $k$: $\vbr{\theta}^a{}_{\mu}$ is real for $k \leq 0$, $\abr{\theta}^a{}_{\mu}$ is real for $k \geq 0$; further, both are real and coincide for $k=0$. It is straightforward to check that, for any choice of $k$, both tetrads yield the real Friedmann-Lemaître-Robertson-Walker (FLRW) metric
\begin{align}
g_{\mu\nu} \dd x^\mu \dd x^\nu = \eta_{ab}\theta^a{}_\mu\theta^b{}_\nu \dd x^\mu \dd x^\nu = - N^2 \dd t^2 + A^2 \left[\frac{\dd r^2}{\chi^2} + r^2 (\dd\vartheta^2 + \sin^2\vartheta \dd\varphi^2) \right]\,.
\end{align}
The torsion building blocks of the field equations can be most easily displayed by introducing $n_\mu$, the components of the homogeneous and isotropic unit normal timelike co-vector field, $n_{\mu} = (-N,0,0,0)$ in $(t,r,\vartheta,\varphi)$ coordinates, and employing equations \eqref{eq:vecT}, \eqref{eq:axT} and \eqref{eq:vecT}. We find for $\vbr{\theta}^a{}_{\mu}$
\begin{align}\label{eq:cosmoTorV}
\vbr{\mathfrak{v}}_{\mu} = 3\left(\frac{i u}{A} - H\right) n_{\mu}\,, \quad
\vbr{\mathfrak{a}}_{\mu} = 0\,, \quad
\vbr{\mathfrak{t}}_{\mu\nu\rho} = 0\,,
\end{align}
while for $\abr{\theta}^a{}_{\mu}$ the result is
\begin{align}\label{eq:cosmoTorA}
\abr{\mathfrak{v}}_{\mu} = -3Hn_{\mu}\,, \quad
\abr{\mathfrak{a}}_{\mu} = -2\frac{u}{A}n_{\mu}\,, \quad
\abr{\mathfrak{t}}_{\mu\nu\rho} = 0\,,
\end{align}
where
\begin{equation}
H = \frac{\partial_tA}{NA}
\end{equation}
is the Hubble parameter. We finally remark that the two different sign choices \(u = \pm\sqrt{k}\) for the axial tetrad constitute inequivalent teleparallel geometries, since the choice of the sign determines the relative orientation of the vector and axial torsion components.

\subsection{The cosmological field equations}\label{ssec:cosmofeq}
Using the tetrad branches~\eqref{eq:tet1} and~\eqref{eq:tet2} given above, we can now derive the cosmological field equations. Since the tetrad field equations turn out to be essentially different for the two tetrad branches, we discuss them in separate sections~\ref{sssec:cosmovec} and~\ref{sssec:cosmoaxi}. The field equations for the axion field take a similar form for both cases, which is therefore jointly discussed in section~\ref{sssec:cosmosca}.

\subsubsection{The vector branch}\label{sssec:cosmovec}
For the vector branch $\vbr{\theta}^a{}_{\mu}$, using the torsion components~\eqref{eq:cosmoTorV} in the field equations~\eqref{eq:tetradfeq}, we find that there are two independent, non-trivial cosmological equations, as a consequence of the cosmological symmetry~\cite{Hohmann:2019nat}. These are the time-time and diagonal space-space components, which are given by
\begin{subequations}\label{eq:cosmoDynV}
\begin{align}
-\frac{9c_v}{A^2}\left(\frac{(\partial_tA)^2}{N^2} + k\right) - \frac{\mathcal{Z}}{N^2}(\partial_t\phi)^2 - 2\kappa^2\mathcal{V} &= 2\kappa^2\rho\,,\\
\frac{3c_v}{A^2N^3}\left[2NA\partial_t^2A + N(\partial_tA)^2 - 2\partial_tNA\partial_tA + kN^3\right] - \frac{\mathcal{Z}}{N^2}(\partial_t\phi)^2 + 2\kappa^2\mathcal{V} &= 2\kappa^2p\,,
\end{align}
\end{subequations}
where we have not fixed a particular time parametrization. One can write a equations in a more compact form by choosing the cosmological time parametrization \(N \equiv 1\), and using the Hubble parameter \(H = \dot{A}/A\), where the dot denotes the derivative with respect to cosmological time. In this case the field equations read
\begin{subequations}
\begin{align}
-9c_v\left(H^2 + \frac{k}{A^2}\right) - \mathcal{Z}\dot{\phi}^2 - 2\kappa^2\mathcal{V} &= 2\kappa^2\rho\,,\\
3c_v\left(2\dot{H} + 3H^2 + \frac{k}{A^2}\right) - \mathcal{Z}\dot{\phi}^2 + 2\kappa^2\mathcal{V} &= 2\kappa^2p\,.
\end{align}
\end{subequations}
Alternatively, we can express them in conformal time, where \(N \equiv A\), as
\begin{subequations}
\begin{align}
-9c_v(\mathcal{H} + k) - \mathcal{Z}\phi'^2 - 2\kappa^2A^2\mathcal{V} &= 2\kappa^2A^2\rho\,,\\
3c_v\left(2\mathcal{H}' + \mathcal{H}^2 + k\right) - \mathcal{Z}\phi'^2 + 2\kappa^2A^2\mathcal{V} &= 2\kappa^2A^2p\,,
\end{align}
\end{subequations}
where the conformal Hubble parameter is \(\mathcal{H} = A'/A\), and the prime denotes the derivative with respect to conformal time. We see that the axion \(\phi\) behave as minimally coupled (pseudo-)scalar field in this branch. This is due to the fact that both the axial and tensor parts of the torsion, which govern the non-minimal coupling in the field equations~\eqref{eq:tetradfeq}, vanish in this case. This is also reflected by the scalar field equation, as we will see below.

\subsubsection{The axial branch}\label{sssec:cosmoaxi}
For the tetrad $\abr{\theta}^a{}_{\mu}$, using the torsion components~\eqref{eq:cosmoTorA}, we again find two independent, non-trivial cosmological equations, which now read
\begin{subequations}\label{eq:cosmoDynA}
\begin{align}
-9c_v\frac{(\partial_tA)^2}{N^2A^2} + 4c_a\frac{k}{A^2} - \frac{\mathcal{Z}}{N^2}(\partial_t\phi)^2 - 2\kappa^2\mathcal{V} &= 2\kappa^2\rho\,,\\
\frac{3c_v}{A^2N^3}\left[2NA\partial_t^2A + N(\partial_tA)^2 - 2\partial_tNA\partial_tA\right] - \frac{4}{3}c_a\frac{k}{A^2} + 2\frac{bu\partial_t\phi}{NA} - \frac{\mathcal{Z}}{N^2}(\partial_t\phi)^2 + 2\kappa^2\mathcal{V} &= 2\kappa^2p\,,
\end{align}
\end{subequations}
using the general time parametrization. In cosmological time, these field equations read
\begin{subequations}
\begin{align}
-9c_vH^2 + 4c_a\frac{k}{A^2} - \mathcal{Z}\dot{\phi}^2 - 2\kappa^2\mathcal{V} &= 2\kappa^2\rho\,,\\
3c_v\left(2\dot{H} + 3H^2\right) - \frac{4}{3}c_a\frac{k}{A^2} + 2\frac{bu\dot{\phi}}{A} - \mathcal{Z}\dot{\phi}^2 + 2\kappa^2\mathcal{V} &= 2\kappa^2p\,,
\end{align}
\end{subequations}
and can equivalently be written in conformal time as
\begin{subequations}
\begin{align}
-9c_v\mathcal{H}^2 + 4c_ak - \mathcal{Z}\phi'^2 - 2\kappa^2A^2\mathcal{V} &= 2\kappa^2A^2\rho\,,\\
3c_v\left(2\mathcal{H}' + \mathcal{H}^2\right) - \frac{4}{3}c_ak + 2bu\phi' - \mathcal{Z}\phi'^2 + 2\kappa^2A^2\mathcal{V} &= 2\kappa^2A^2p\,,
\end{align}
\end{subequations}
These equations differ from the result for the vector branch in two important aspect. First, note that in the axial case~\eqref{eq:cosmoDynA} the parameter \(c_a\) governs the contribution from the non-vanishing curvature parameter \(k\), while in the vector case~\eqref{eq:cosmoDynV} this contribution is subsumed under the contribution from the Hubble parameter with a common coefficient \(c_v\). For the TEGR values~\eqref{eq:tegr} of these constants, this distinction vanishes, and \(k\) contributes equally for both tetrad branches. This has been previously noted in the application of the two cosmologically symmetric tetrad branches to new general relativity~\cite{Hohmann:2020zre}.

The second important observation is the appearance of a new term \(\sim bu\partial_t\phi\), which is not present in the vector branch. Under a parity transformation, which affects both the axion \(\phi \mapsto -\phi\) and the tetrad \(u \mapsto -u\), this term is invariant. It follows from the explicit dependence on \(u\) that the two tetrads for \(u = \pm\sqrt{k}\), which yield the same curvature parameter \(k = u^2\), and hence the same FLRW metrics, but are distinguished by a parity transformation, attain different dynamics if \(\phi\) possesses a non-vanishing time dependence. This feature of branching dynamics in the axial branch of cosmologically symmetric tetrads is specific to theories in which a non-minimal coupling to the axial torsion exists.

\subsubsection{Scalar field equations}\label{sssec:cosmosca}
We finally come to the cosmological field equation for the axion field. This can be written uniformly for both tetrad branches as
\begin{equation}\label{eq:cosmoDynS}
2\frac{\mathcal{Z}}{N^2}\left(\partial_t^2\phi - \frac{\partial_tN\partial_t\phi}{N} + 3\frac{\partial_tA\partial_t\phi}{A}\right) + \frac{1}{N^2}\mathcal{Z}_{\phi}(\partial_t\phi)^2 + 2\kappa^2\mathcal{V}_{\phi} + bP = 0\,,
\end{equation}
where we used the general time parametrization. In cosmological time, this simplifies to
\begin{equation}
2\mathcal{Z}\left(\ddot{\phi} + 3H\dot{\phi}\right) + \mathcal{Z}_{\phi}\dot{\phi}^2 + 2\kappa^2\mathcal{V}_{\phi} + bP = 0\,,
\end{equation}
or equivalently, in conformal time reads
\begin{equation}
2\mathcal{Z}\left(\phi'' + 2\mathcal{H}\phi'\right) + \mathcal{Z}_{\phi}\phi'^2 + 2\kappa^2A^2\mathcal{V}_{\phi} + bPA^2 = 0\,.
\end{equation}
A particular role is attributed to the term \(bP\). For the two different tetrad branches, this term takes the values
\begin{equation}
bP = b\mathfrak{v}_{\mu}\mathfrak{a}^{\mu} = \begin{cases}
-6bu\frac{\partial_tA}{NA^2} & \text{for the axial tetrad } \abr{\theta}^a{}_{\mu}\,,\\
0 & \text{for the vector tetrad } \vbr{\theta}^a{}_{\mu}\,.
\end{cases}
\end{equation}
This reflects our previous findings from the tetrad equations, which show that for the vector branch the axion field becomes minimally coupled, while for the axial branch one obtains a non-minimal coupling.

\subsection{Analysing the field equations}\label{ssec:cosmoana}
We finally compare the cosmological dynamics we obtained to the Friedmann equations. Choosing the TEGR values~\eqref{eq:tegr} in the action, we can write the general field equations in the form of the Einstein equations
\begin{equation}
R_{\mu\nu} - \frac{1}{2}g_{\mu\nu}R = \kappa^2\Theta^{\text{eff}}_{\mu\nu}\,,
\end{equation}
where \(\Theta^{\text{eff}}_{\mu\nu}\) is the effective energy-momentum tensor constituted by the matter energy-momentum, as well as the contributions from the axion field. In the case of cosmological symmetry, and choosing the cosmological time parametrization \(N \equiv 1\) for convenience, we thus obtain the effective Friedmann equations
\begin{subequations}
\begin{align}
3\left(H^2 + \frac{k}{A^2}\right) & = \kappa^2(\rho + \rho_{\phi})\label{eq:friedcons}\,,\\
-\left(2\dot{H} + 3H^2 + \frac{k}{A^2}\right) &= \kappa^2(p + p_{\phi})\,,
\end{align}
\end{subequations}
where the effective energy density and pressure of the axion field are given by
\begin{equation}
\rho_{\phi} = \frac{1}{2\kappa^2}\mathcal{Z}\dot{\phi}^2 + \mathcal{V}\,, \quad
p_{\phi} = \frac{1}{2\kappa^2}\mathcal{Z}\dot{\phi}^2 - \mathcal{V} - \begin{cases}
\frac{bu\dot{\phi}}{\kappa^2A} & \text{for the axial tetrad } \abr{\theta}^a{}_{\mu}\,,\\
0 & \text{for the vector tetrad } \vbr{\theta}^a{}_{\mu}\,.
\end{cases}
\end{equation}
Since for the vector branch \(\vbr{\theta}^a{}_{\mu}\) the axion becomes minimally coupled, in the following we will discuss only the axial branch \(\abr{\theta}^a{}_{\mu}\). Due to its dependence \(\sim u/A\) on the spatial curvature, we see that the contribution from the axion coupling becomes negligible in the late universe. Hence, one may expect effects on the background dynamics only in the pre-inflation era. Further, the explicit presence of \(u = \pm\sqrt{k}\) signals that the contribution of this term breaks parity invariance, so that the two tetrads \(\abr{\theta}^a{}_{\mu}\) corresponding to these different values of \(u\) attain different dynamics, despite inducing the same FLRW metric. To qualitatively study such effects, one may neglect the matter contribution, hence assuming a vacuum solution \(\rho = p = 0\), and introduce a parametrization \((\alpha,\beta)\) of the phase space by setting
\begin{equation}
\dot{\phi} = \sqrt{2\kappa^2\frac{\mathcal{V}}{\mathcal{Z}}}\frac{\alpha}{\sqrt{1 - \alpha^2}}\,, \quad
H = \sqrt{\frac{\kappa^2}{3 - 3\alpha^2}\mathcal{V}}\cos\beta\,, \quad
A = u\left(\sqrt{\frac{\kappa^2}{3 - 3\alpha^2}\mathcal{V}}\sin\beta\right)^{-1}\,,
\end{equation}
which identically solves the Friedmann constraint~\eqref{eq:friedcons}. These new variables are constrained by the conditions
\begin{equation}
-1 < \alpha < 1\,, \quad
\sgn u = \sgn\sin\beta\,,
\end{equation}
and so we can study the effect of the choice of \(u\) by considering either \(0 < \beta < \pi\) or \(-\pi < \beta < 0\). To further simplify the cosmological dynamics, we consider a constant roll approximation by choosing a constant potential \(\mathcal{V} = \Lambda/\kappa^2\) corresponding to a cosmological constant \(\Lambda\) and canonical kinetic coupling \(\mathcal{Z} = 1\). In this case the dynamics is described by the autonomous dynamical system
\begin{equation}\label{eq:dynsym}
\frac{\dot{\alpha}}{\sqrt{\Lambda}} = \left(\frac{b}{\sqrt{2}}\sin\beta - \sqrt{3}\alpha\right)\sqrt{1 - \alpha^2}\cos\beta\,, \quad
\frac{\dot{\beta}}{\sqrt{\Lambda}} = \left(\frac{3\alpha^2 - 1}{\sqrt{3}} - \frac{b}{\sqrt{2}}\alpha\sin\beta\right)\frac{\sin\beta}{\sqrt{1 - \alpha^2}}\,.
\end{equation}
We see that for \(b = 0\), this system is indeed invariant under the coordinate transformation \(\beta \mapsto -\beta\), while this invariance is broken for non-vanishing \(b\), in concordance with the breaking of parity invariance. Note that this can be compensated by simultaneously reversing the sign of \(\dot{\phi}\), and so the system is invariant under the coordinate transformation \((\alpha,\beta) \mapsto (-\alpha,-\beta)\). Hence, for every solution with \(u = \sqrt{k}\), there also exists a solution for \(u = -\sqrt{k}\) and the opposite sign of \(\dot{\phi}\).

\begin{figure}[htbp]
\centering
\subfigure[$b = 0$]{\includegraphics[width=0.48\textwidth]{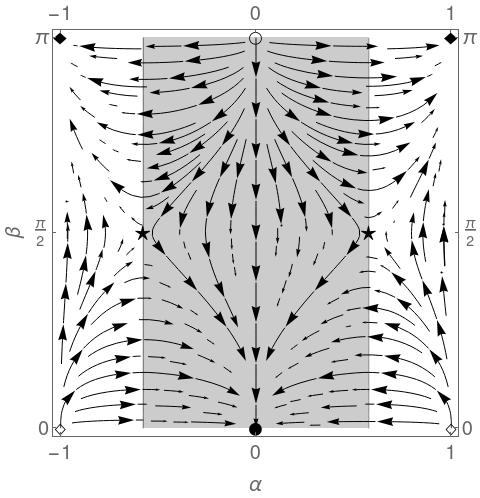}\label{fig:min}}
\subfigure[$0 < b < \sqrt{8/3}$]{\includegraphics[width=0.48\textwidth]{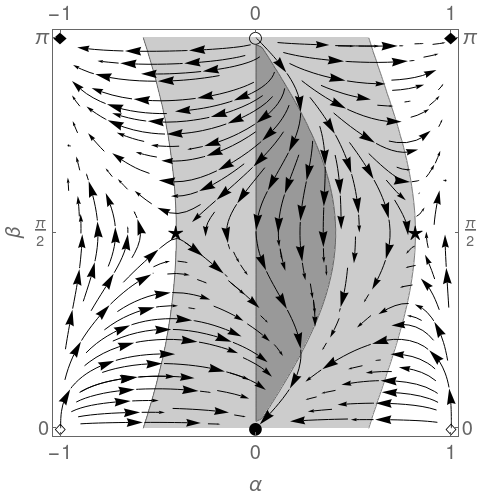}\label{fig:weak}}
\subfigure[$b = \sqrt{8/3}$]{\includegraphics[width=0.48\textwidth]{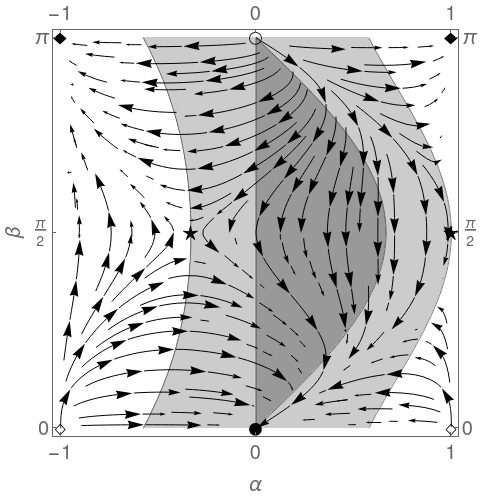}\label{fig:crit}}
\subfigure[$b > \sqrt{8/3}$]{\includegraphics[width=0.48\textwidth]{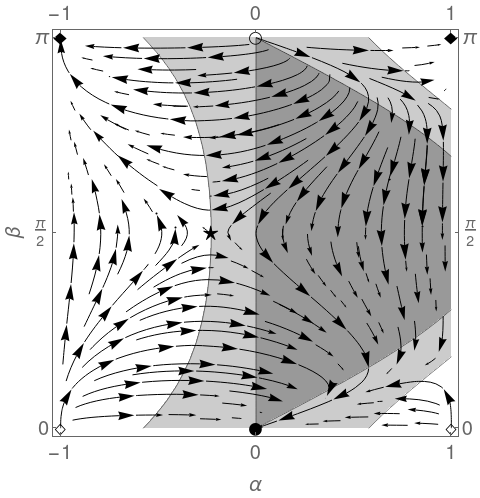}\label{fig:strong}}
\caption{Qualitative phase diagrams for the dynamical system~\eqref{eq:dynsym} for different values of \(b\). Only the range \(0 < \beta < \pi\) for \(u > 0\) is shown; one obtains \(-\pi < \beta < 0\) by rotating 180 around \((0,0)\). Filled / open circles denote stable (expanding) / unstable (contracting) de Sitter fixed points. Filled / empty diamonds denote Big Crunch / Big Bang. Stars denote saddle fixed points for \(\alpha \neq 0\). Shades from dark to white indicate phantom behavior \(w < -1\), acceleration \(-1 < w < -1/3\) and deceleration \(w > -1/3\), where \(w = p_{\phi}/\rho_{\phi}\). Note the asymmetry under reflection \(\alpha \mapsto -\alpha\) for \(b \neq 0\), which is equivalent to an asymmetry under \(\beta \mapsto -\beta\).}
\label{fig:phasep}
\end{figure}

It is instructive to consider the fixed points of this system. From the first equation we see that \(\dot{\alpha} = 0\) if any of the conditions
\begin{equation}
\alpha = \pm 1\,, \quad
\alpha = \frac{b}{\sqrt{6}}\sin\beta\,, \quad
\frac{\beta}{\pi} - \frac{1}{2} \in \mathbb{Z}
\end{equation}
is satisfied. The first condition lies on the boundary \(A \to 0\) of the phase space corresponding to a Big Bang / Big Crunch singularity. For the second condition, the term in parentheses in the second equation simplifies and one obtains
\begin{equation}
\frac{\dot{\beta}}{\sqrt{\Lambda}} = \frac{\sin\beta}{\sqrt{3 - 3\alpha^2}}\,,
\end{equation}
which vanishes only for \(\beta/\pi \in \mathbb{Z}\), and then implies \(\alpha = 0\); these fixed points correspond to an infinitely expanding or contracting universe with \(A \to \infty\) and \(H \to \pm\sqrt{\Lambda/3}\), where the effect of the axion coupling becomes negligible. Hence, we focus on the remaining condition, which corresponds to \(\sin\beta = \pm 1 = \sgn u\). In this case we find \(\dot{\beta} = 0\) for
\begin{equation}
\alpha^2 - \frac{b}{\sqrt{6}}\alpha\sgn u - \frac{1}{3} = 0 \quad \Rightarrow \quad \alpha = \frac{b \pm \sqrt{b^2 + 8}}{2\sqrt{6}}\sgn u\,.
\end{equation}
Note that if the upper, positive sign is chosen in the formula above, then the fixed point lies inside the phase space \(|\alpha| < 1\) only for \(b < \sqrt{8/3}\), while for the negative sign one must demand \(b > -\sqrt{8/3}\). Hence, if \(|b| < \sqrt{8/3}\) one obtains two fixed points in the phase space, and only one for \(|b| > \sqrt{8/3}\). The qualitative behavior for these different cases is shown in the phase portraits in figure~\ref{fig:phasep}, where the last figure~\ref{fig:strong} shows that one of the two saddle points disappears, so that all trajectories emanating from the Big Bang at \((1,0)\) reach the de Sitter fixed point at \((0,0)\) instead of the Big Crunch at \((1,\pi)\). A full quantitative analysis, taking into account also matter and a potential driving inflation, using the method of dynamical systems~\cite{Bahamonde:2017ize}, would exceed the scope of this article.

We finally remark that although the axion enters the cosmological background equations only for the axial tetrad \(\abr{\theta}^a{}_{\mu}\), it contributes to perturbations around all cosmological background tetrads, including the spatially flat tetrad obtained for \(k = 0\) at the intersection of the axial and vector branches~\cite{Li:2020xjt}. However, here we limit our discussion to the background dynamics of the homogeneous and isotropic FLRW spacetime, and leave the study of cosmological perturbations for future work.

\section{Extensions and alternatives}\label{sec:ext}
In the previous section we have considered a single axion field with a constant coupling to teleparallel gravity. We now extend this most simple teleparallel axial model, and discuss several generalizations. The aim of this section is to give an overview of possible axion couplings in more general teleparallel gravity theories, without studying their dynamics in detail, which would exceed the scope of this article. Further, we focus on the gravitational part of the action only, since our aim is to discuss the possible axion coupling terms; possible matter couplings to complement these actions are given in~\cite{BeltranJimenez:2020sih}, and the most general contribution compatible with the cosmological symmetry is the hyperfluid discussed in~\cite{Iosifidis:2021nra}. In particular, in section~\ref{ssec:multi}, we consider multiple axion fields. Dynamical couplings are discussed in section~\ref{ssec:dynamical}. Finally, in section~\ref{ssec:symtele} we present an alternative approach to couple axions to symmetric teleparallel gravity instead of the torsional geometry we considered here. An outlook towards general teleparallel gravity theories featuring both torsion and nonmetricity is given in section~\ref{ssec:gentele}.

\subsection{Multiple axions}\label{ssec:multi}
In the action~\eqref{eq:gravact} we considered a single axion field \(\phi\) which is non-minimally coupled to the parity-odd torsion invariants. A natural generalization of this model is to replace this single field by a multiplet \(\boldsymbol{\phi} = (\phi^A, A = 1, \ldots, n)\) of \(n\) pseudo-scalar fields, and allow each of these fields to couple with an independent coupling pair \(b_A\) and \(\tilde{b}_A\) of coupling parameters to the pseudo-scalar invariants. Also the parameter functions \(\mathcal{V}\) and \(\mathcal{Z}\) receive generalizations compared to the single-field case. Both of them become functions of all pseudo-scalar fields \(\phi^A\). Further, the single kinetic coupling function is replaced by an indexed quantity \(\mathcal{Z}_{AB}\), which is symmetric in both indices. Hence, the total action takes the form
\begin{equation}\label{eq:multigravact}
S_{\text{g}}[\theta, \omega, \phi] = \int\dd^4x\,\theta\,\frac{1}{2\kappa^2}\left[c_vT_{\text{vec}} + c_aT_{\text{axi}} + c_tT_{\text{ten}} + b_A\phi^AP + \tilde{b}_A\phi^A\tilde{P} + \mathcal{Z}_{AB}(\boldsymbol{\phi})g^{\mu\nu}\partial_{\mu}\phi^A\partial_{\nu}\phi^B + 2\kappa^2\mathcal{V}(\boldsymbol{\phi})\right]\,.
\end{equation}
It is instructive to study how the field equations change for this generalized theory. For the tetrad equations~\eqref{eq:tetradfeq}, one obtains the straightforward generalization
\begin{equation}
\begin{split}
2\kappa^2\Theta_{\mu\nu} &= c_a\left[\mathfrak{a}^{\rho}\mathfrak{a}_{(\rho}g_{\mu\nu)} - \frac{8}{9}\epsilon_{\nu\alpha\beta\gamma}\mathfrak{a}^{\alpha}\mathfrak{t}_{\mu}{}^{\beta\gamma} - \frac{4}{9}\epsilon_{\mu\nu\rho\sigma}\mathfrak{a}^{\rho}\mathfrak{v}^{\sigma} - \frac{2}{3}\epsilon_{\mu\nu\rho\sigma}\lc{\nabla}^{\rho}\mathfrak{a}^{\sigma}\right]\\
&\phantom{=}+ c_v\left[\mathfrak{v}^{\rho}\mathfrak{v}_{(\rho}g_{\mu\nu)} + \frac{8}{3}\mathfrak{t}_{\mu[\rho\nu]}\mathfrak{v}^{\rho} + 4g_{\mu[\nu}\lc{\nabla}^{\rho}\mathfrak{v}_{\rho]} - \epsilon_{\mu\nu\rho\sigma}\mathfrak{a}^{\rho}\mathfrak{v}^{\sigma}\right]\\
&\phantom{=}+ c_t\left[\frac{4}{3}\mathfrak{t}_{\alpha[\beta\gamma]}\mathfrak{t}^{\alpha\beta\gamma}g_{\mu\nu} - \frac{8}{3}\mathfrak{t}_{\mu[\rho\sigma]}\mathfrak{t}_{\nu}{}^{\rho\sigma} + 4\lc{\nabla}^{\rho}\mathfrak{t}_{\mu[\nu\rho]} - \frac{4}{3}\mathfrak{t}_{\nu[\mu\rho]}\mathfrak{v}^{\rho} + \epsilon_{\mu\alpha\beta\gamma}\mathfrak{a}^{\alpha}\mathfrak{t}_{\nu}{}^{\beta\gamma}\right]\\
&\phantom{=}+ b_A\phi^A\left[\mathfrak{a}^{\rho}\mathfrak{v}_{(\rho}g_{\mu\nu)} - \frac{4}{3}\mathfrak{t}_{\mu[\nu\rho]}\mathfrak{a}^{\rho} - \frac{4}{9}\epsilon_{\nu\alpha\beta\gamma}\mathfrak{v}^{\alpha}\mathfrak{t}_{\mu}{}^{\beta\gamma}\right] + b_A\left[2g_{\mu[\nu}\lc{\nabla}^{\rho}(\phi^A\mathfrak{a}_{\rho]}) - \frac{1}{3}\epsilon_{\mu\nu\rho\sigma}\lc{\nabla}^{\rho}(\phi^A\mathfrak{v}^{\sigma})\right]\\
&\phantom{=}+ \tilde{b}_A\phi^A\left[6g_{\nu[\mu}\epsilon_{\alpha\beta]\gamma\delta}\mathfrak{t}^{\rho\alpha\beta}\mathfrak{t}_{\rho}{}^{\gamma\delta} - \frac{8}{3}\mathfrak{t}_{\mu}{}^{[\alpha\beta]}\mathfrak{t}_{(\nu}{}^{\gamma\delta}\epsilon_{\alpha)\beta\gamma\delta} - \frac{4}{3}g_{\nu\rho}g_{\mu[\sigma}\mathfrak{v}_{\alpha]}\mathfrak{t}^{(\rho}{}_{\beta\gamma}\epsilon^{\sigma)\alpha\beta\gamma} - 3\mathfrak{t}_{\nu[\mu\rho]}\mathfrak{a}^{\rho}\right]\\
&\phantom{=}- 4\tilde{b}_Ag_{\mu\rho}g_{\nu[\sigma}\epsilon^{\alpha\beta\gamma(\rho}\lc{\nabla}_{\alpha]}(\phi^A\mathfrak{t}^{\sigma)}{}_{\beta\gamma}) - 2\mathcal{Z}_{AB}\partial_{\mu}\phi^A\partial_{\nu}\phi^B + \mathcal{Z}_{AB}\partial_{\rho}\phi^A\partial_{\sigma}\phi^Bg^{\rho\sigma}g_{\mu\nu} + 2\kappa^2\mathcal{V}g_{\mu\nu}\,.
\end{split}
\end{equation}
By variation with respect to the axion fields one obtains the generalization of the scalar field equation~\eqref{eq:scalarfeq} as
\begin{equation}
0 = \mathcal{Z}_{AB}\lc{\square}\phi^B + g^{\mu\nu}\left(\partial_C\mathcal{Z}_{AB}\partial_{\mu}\phi^C\partial_{\nu}\phi^B - \frac{1}{2}\partial_A\mathcal{Z}_{BC}\partial_{\mu}\phi^B\partial_{\nu}\phi^C\right) - \kappa^2\partial_A\mathcal{V} - b_AP - \tilde{b}_A\tilde{P}\,,
\end{equation}
where the term which is quadratic in the first order derivatives splits into two terms, as is known from coupling multiple scalar fields to curvature~\cite{Damour:1992we} or torsion~\cite{Hohmann:2018rwf,Hohmann:2018ijr}.

A case of particular interest is obtained for two scalar fields \(\boldsymbol{\phi} = (\phi, \tilde{\phi})\), together with the couplings \((b_1, b_2) = (b_1, 0)\) and \((\tilde{b}_1, \tilde{b}_2) = (0, \tilde{b}_2)\), so that each scalar field is coupled to exactly one of the parity-odd invariants \(P\) and \(\tilde{P}\). It follows that in cosmological symmetry, only one field enters the cosmological field equations~\eqref{eq:cosmoDynA} for the axial tetrad with a non-trivial contribution in addition to the minimal coupling term, while the other field appears minimally coupled in this equation. However, note that in general both fields are still coupled via the kinetic coupling \(\mathcal{Z}_{AB}\) in their field equation~\eqref{eq:cosmoDynS}.

\subsection{Dynamical couplings}\label{ssec:dynamical}
In the teleparallel axion models considered thus far we have assumed that the three parity-even torsion invariants \(T_{\text{vec}}, T_{\text{axi}}, T_{\text{ten}}\) enter the action with constant coefficients \(c_v, c_a, c_t\), and that the coupling of the axion to the parity-odd terms \(P\) and \(\tilde{P}\) is likewise governed by constants \(b\) and \(\tilde{b}\). A straightforward generalization of this assumptions is to allow for these coefficients to depend on the value of the pseudo-scalar field \(\phi\) itself, i.e., to replace them by functions \(\mathcal{C}_v, \mathcal{C}_a, \mathcal{C}_t, \mathcal{B}, \tilde{\mathcal{B}}\) of \(\phi\). The action then takes the form
\begin{equation}\label{eq:dyngravact}
S_{\text{g}}[\theta, \omega, \phi] = \int\dd^4x\,\theta\,\frac{1}{2\kappa^2}\left[\mathcal{C}_v(\phi)T_{\text{vec}} + \mathcal{C}_a(\phi)T_{\text{axi}} + \mathcal{C}_t(\phi)T_{\text{ten}} + \mathcal{B}(\phi)\phi P + \tilde{\mathcal{B}}(\phi)\phi\tilde{P} + \mathcal{Z}(\phi)g^{\mu\nu}\partial_{\mu}\phi\partial_{\nu}\phi + 2\kappa^2\mathcal{V}(\phi)\right]\,.
\end{equation}
In addition, one may apply the previously mentioned generalization to multiple axion fields. This yields an action of the form
\begin{multline}\label{eq:dynmultigravact}
S_{\text{g}}[\theta, \omega, \phi] = \int\dd^4x\,\theta\,\frac{1}{2\kappa^2}\Big[\mathcal{C}_v(\boldsymbol{\phi})T_{\text{vec}} + \mathcal{C}_a(\boldsymbol{\phi})T_{\text{axi}} + \mathcal{C}_t(\boldsymbol{\phi})T_{\text{ten}}\\
+ \mathcal{B}_A(\boldsymbol{\phi})\phi^AP + \tilde{\mathcal{B}}_A(\boldsymbol{\phi})\phi^A\tilde{P} + \mathcal{Z}_{AB}(\boldsymbol{\phi})g^{\mu\nu}\partial_{\mu}\phi^A\partial_{\nu}\phi^B + 2\kappa^2\mathcal{V}(\boldsymbol{\phi})\Big]\,,
\end{multline}
where the parameter functions depend on all fields constituting the multiplet \(\boldsymbol{\phi}\). We omit the field equations here for brevity, but remark that they are straightforward to derive.

\subsection{Symmetric teleparallel axions}\label{ssec:symtele}
In this article we have made use of the teleparallel geometry, whose connection satisfies the condition~\eqref{eq:telep} of flat and metric compatibility. Another class of theories is known as symmetric teleparallel gravity~\cite{Nester:1998mp}. In these models the fundamental field variables are a metric \(g_{\mu\nu}\) and an affine connection \(\Gamma^{\mu}{}_{\nu\rho}\), which are constrained such that the corresponding metric-affine geometry has vanishing curvature \(R^{\mu}{}_{\nu\rho\sigma}\) and torsion \(T^{\mu}{}_{\nu\rho}\), but nonvanishing nonmetricity \(Q_{\rho\mu\nu}\). Since the latter is symmetric in its last two indices, it allows the construction of five quadratic scalar invariants
\begin{equation}\label{eq:qseven}
\mathcal{Q}_1 = Q^{\rho\mu\nu}Q_{\rho\mu\nu}\,, \quad
\mathcal{Q}_2 = Q^{\mu\nu\rho}Q_{\rho\mu\nu}\,, \quad
\mathcal{Q}_3 = Q^{\rho\mu}{}_{\mu}Q_{\rho\nu}{}^{\nu}\,, \quad
\mathcal{Q}_4 = Q^{\mu}{}_{\mu\rho}Q_{\nu}{}^{\nu\rho}\,, \quad
\mathcal{Q}_5 = Q^{\mu}{}_{\mu\rho}Q^{\rho\nu}{}_{\nu}\,,
\end{equation}
as well as one pseudo-scalar
\begin{equation}\label{eq:qsodd}
\hat{\mathcal{Q}} = \epsilon^{\mu\nu\rho\sigma}Q_{\mu\nu\lambda}Q_{\rho\sigma}{}^{\lambda}\,.
\end{equation}
These terms can also be expressed in an irreducible decomposition on the nonmetricity tensor~\cite{Hehl:1994ue}. Hence, one can proceed in full analogy to the approach we presented in this article and couple a pseudo-scalar field to this parity-odd term. A possible action one can conceive in analogy to the teleparallel axion model~\eqref{eq:gravact} as generalization of newer general relativity~\cite{BeltranJimenez:2017tkd,BeltranJimenez:2018vdo,Koivisto:2019jra} is then given by
\begin{equation}\label{eq:symteleaction}
S_{\text{g}}[g, \Gamma, \phi] = \int\dd^4x\,\sqrt{-g}\,\frac{1}{2\kappa^2}\left[\sum_{i = 1}^5c_i\mathcal{Q}_i + b\phi\hat{\mathcal{Q}} + \mathcal{Z}(\phi)g^{\mu\nu}\partial_{\mu}\phi\partial_{\nu}\phi + 2\kappa^2\mathcal{V}(\phi)\right]\,.
\end{equation}
In the case that the parameters take the values
\begin{equation}
c_1 = -\frac{1}{4}\,, \quad
c_2 = \frac{1}{2}\,, \quad
c_3 = \frac{1}{4}\,, \quad
c_4 = 0\,, \quad
c_5 = -\frac{1}{2}
\end{equation}
one obtains the symmetric teleparallel equivalent of relativity~\cite{Nester:1998mp} (STEGR), together with \emph{symmetric teleparallel axion}, which is non-minimally coupled to the unique parity-odd invariant of non-metricity. Hence, also this geometric framework allows for a canonical method to couple axions to general relativity.

It is now obvious that one can generalize the symmetric teleparallel action~\eqref{eq:symteleaction} in the same way as in the action~\eqref{eq:gravact} for the torsional model. A multiplet \(\boldsymbol{\phi}\) can be accommodated in the form
\begin{equation}
S_{\text{g}}[g, \Gamma, \phi] = \int\dd^4x\,\sqrt{-g}\,\frac{1}{2\kappa^2}\left[\sum_{i = 1}^5c_i\mathcal{Q}_i + b_A\phi^A\hat{\mathcal{Q}} + \mathcal{Z}_{AB}(\boldsymbol{\phi})g^{\mu\nu}\partial_{\mu}\phi^A\partial_{\nu}\phi^B + 2\kappa^2\mathcal{V}(\boldsymbol{\phi})\right]\,,
\end{equation}
with coupling constants \(b_A\) for each axion field, in analogy to the action~\eqref{eq:multigravact}. Dynamical couplings may be introduced by studying the action
\begin{equation}
S_{\text{g}}[g, \Gamma, \phi] = \int\dd^4x\,\sqrt{-g}\,\frac{1}{2\kappa^2}\left[\sum_{i = 1}^5\mathcal{C}_i(\phi)\mathcal{Q}_i + \mathcal{B}(\phi)\phi\hat{\mathcal{Q}} + \mathcal{Z}(\phi)g^{\mu\nu}\partial_{\mu}\phi\partial_{\nu}\phi + 2\kappa^2\mathcal{V}(\phi)\right]\,,
\end{equation}
in the single field case in analogy to the action~\eqref{eq:dyngravact}, or finally
\begin{equation}
S_{\text{g}}[g, \Gamma, \phi] = \int\dd^4x\,\sqrt{-g}\,\frac{1}{2\kappa^2}\left[\sum_{i = 1}^5\mathcal{C}_i(\boldsymbol{\phi})\mathcal{Q}_i + \mathcal{B}(\boldsymbol{\phi})\phi\hat{\mathcal{Q}} + \mathcal{Z}_{AB}(\boldsymbol{\phi})g^{\mu\nu}\partial_{\mu}\phi^A\partial_{\nu}\phi^B + 2\kappa^2\mathcal{V}(\boldsymbol{\phi})\right]\,,
\end{equation}
allowing for multiple axions as in the action~\eqref{eq:dynmultigravact}. We will not pursue these extensions further in this article, and leave their investigation for future work. We also remark that a different class of scalar field couplings to more general parity-odd invariants, which also include higher-order derivatives, were studied in~\cite{Conroy:2019ibo}.

\subsection{General teleparallel axions}\label{ssec:gentele}
In the previous sections we have studied geometries in which either torsion or nonmetricity appears as the only characteristic tensor field, which enters into the action as the mediator of the gravitational interaction. Recently a class of general teleparallel theories has been proposed, in which both torsion and nonmetricity are present, while curvature is still assumed to vanish~\cite{Jimenez:2019ghw}. The action for these theories is taken as a linear combination of the three parity-even torsion invariants~\eqref{eq:tseven}, the five parity-even nonmetricity invariants~\eqref{eq:qseven} and three parity-even cross terms
\begin{equation}
Q_{\mu\nu\rho}T^{\rho\mu\nu}\,, \quad
Q^{\mu}{}_{\mu\rho}T_{\nu}{}^{\nu\rho}\,, \quad
Q_{\rho\mu}{}^{\mu}T_{\nu}{}^{\nu\rho}\,.
\end{equation}
It follows that also these general teleparallel gravity theories allow for a coupling of axion fields. In addition to the two parity-odd torsion invariants~\eqref{eq:tsodd} and the parity-odd nonmetricity invariant~\eqref{eq:qsodd}, there are three parity-odd invariants
\begin{equation}
\epsilon^{\mu\nu\rho\sigma}Q_{\mu\nu}{^\tau}T_{(\tau\rho)\sigma}\,, \quad
\epsilon^{\mu\nu\rho\sigma}Q^{\tau}{}_{\tau\mu}T_{\nu\rho\sigma}\,, \quad
\epsilon^{\mu\nu\rho\sigma}Q_{\mu\tau}{}^{\tau}T_{\nu\rho\sigma}\,,
\end{equation}
combining both tensor fields, so that the most general teleparallel axion would feature six coupling terms in its gravitational action.

\section{Conclusion}\label{sec:conclusion}
We have proposed a new class of teleparallel gravity theories featuring a pseudo-scalar field with two independent couplings to the two pseudo-scalar torsion invariants, in order to obtain the most general class of teleparallel gravity theories which are parity-invariant and quadratic in the torsion tensor, without additional derivative terms. This model is a natural generalization of the model presented in~\cite{Li:2020xjt,Chatzistavrakidis:2020wum}, where only a single coupling to the Nieh-Yan term~\cite{Nieh:1981ww} is considered. In analogy to the QCD and electrodynamics nomenclature, we call this pseudo-scalar field \emph{teleparallel axion}. Furthermore we followed the same line of argument and discussed several possible extensions to other geometries, giving rise to what we called \emph{symmetric teleparallel axion} (coupling to the parity odd quadratic non-metricity scalar), and \emph{general teleparallel axion} (coupling in addition to the three parity odd torsion-non-metricity scalars, in case one considers a connection with torsion and non-metricity).

As an application, we derived the cosmological field equations for a homogeneous and isotropic FLRW spacetime, which is implemented by the two branches constituting the most general homogeneous and isotropic teleparallel geometries~\cite{Hohmann:2020zre}. We find that for one branch, which is called the vector branch, the axion contributes to the background field equations as a minimally coupled (pseudo-)scalar field. For the other branch, called the axion branch, we find a non-vanishing contribution from the non-minimal coupling to the parity-odd torsion term in the action. Further, we find that the sign of this contribution depends on the choice of a sign in the axial tetrad, which determines the relative orientation of vector and axial torsion. We studied the quantitative behavior of this contribution for a simple model of an axion-dominated cosmology and found the appearance of new saddle fixed points from the axion coupling.

In this work we have considered only homogeneous and isotropic cosmology, which is described by an exact FLRW geometry. This can be considered as a starting point for further studies. The most straightforward is to apply the method of dynamical systems~\cite{Bahamonde:2017ize} and study the axion contribution in the light of inflation. As another example, one may consider more general cosmological models, such as Bianchi spacetimes, or the influence of the matter sector on the combined dynamics, which may in general also include a hyperfluid coupled to the flat connection~\cite{Iosifidis:2021nra}. Further, one may consider perturbations of the teleparallel geometry around the homogeneous and isotropic tetrads we used here. It has already been found for a simpler teleparallel axion model that a non-vanishing axion background breaks the parity symmetry in the propagation of gravitational waves~\cite{Li:2020xjt}. Further effects may be expected for perturbations around the spatially curved FLRW tetrads we have used in this article~\cite{Hohmann:2020vcv}.

Another possibility to extend our studies is to generalize the notion of tetrads and spin connections to become complex, provided that the metric remains real. This generalization would then allow studying the axial branch of cosmologically symmetric tetrads also in the case \(k < 0\) of an open universe~\cite{Capozziello:2018hly,Hohmann:2019nat,Hohmann:2020zre}. However, the torsion tensor, from which the teleparallel gravity action is constructed, would become complex for such generalized, complex tetrads. Hence, this would also require including additional terms in the action to cancel the imaginary part. We will not pursue such extensions here and leave them for further studies.

\begin{acknowledgments}
The authors acknowledge support by the Estonian Ministry for Education and Science through the Personal Research Funding Grants PRG356 and PSG489, as well as the European Regional Development Fund through the Center of Excellence TK133 ``The Dark Side of the Universe''. C.P. was also funded by the Deutsche Forschungsgemeinschaft (DFG, German Research Foundation) - Project Number 420243324.
\end{acknowledgments}

\bibliography{ngraxcosmo}
\end{document}